\documentclass[aps,prd]{revtex4}
\usepackage{epsfig}

\begin{document}
\title{Influence of gauge boson mass on the staggered spin susceptibility}
\author{Jian-Feng Li$^{1,7}$, Feng-Yao Hou$^{2,7}$, Zhu-Fang Cui$^{3,7}$, Hong-Tao Feng$^{4,7}$, Yu Jiang $^{5,7}$, Hong-Shi Zong$^{3,6,7}$}
\address{$^{1}$ College of  Mathematics and Physics, Nantong University, Nantong 226019, China}
\address{$^{2}$ Institute of Theoretical Physics, CAS, Beijing 100190, China}
\address{$^{3}$ Department of Physics, Nanjing University, Nanjing 210093, China}
\address{$^{4}$ Department of Physics, Southeast University, Nanjing 211198, China}
\address{$^{5}$ College of Mathematics, Physics and Information Engineering, Zhejiang Normal University, Jinhua 321004, China}
\address{$^{6}$ Joint Center for Particle, Nuclear Physics and Cosmology, Nanjing 210093, China}
\address{$^{7}$ State Key Laboratory of Theoretical Physics, Institute of Theoretical Physics, CAS, Beijing 100190, China}

\begin{abstract}
Based on the study of the linear response of the fermion propagator in the presence of an external scalar field, we calculate  the staggered spin susceptibility in the low energy limit in the framework of the Dyson-Schwinger approach. We analyze the effect of a finite gauge boson mass on the staggered spin susceptibility in both Nambu phase and Wigner phase. It is found that the gauge boson mass suppresses the staggered spin susceptibility in
Wigner phase. In addition, we try to give an explanation for why the antiferromagnetic spin correlation increases when the doping is lowered.

\bigskip

\end{abstract}

\maketitle
\section{introduction}
Quantum electrodynamics in (2+1) dimensions (QED$_3$) has attracted much interest over the past few years. It has many features similar to Quantum Chromodynamics (QCD), such as dynamical chiral symmetry breaking in the chiral limit and confinement \cite{a1,a2,a3,a4,a5,a6,a7,a8,a9,a10,a11,b1,a12}. Moreover, it is super-renormalizable, so it does not suffer from the ultraviolet divergence which are present in QED$_4$. Because of these reasons it can serve as a toy model of QCD. In parallel with its relevance as a tool through which to develop insight into aspects of QCD, QED$_3$ is also found to be equivalent to the low-energy effective theories of strongly correlated electronic systems. Recently, QED$_3$ has been proved to be a useful tool to study antiferromagnetic spin correlation in the so-called staggered flux liquid phase in high T$_c$ cuprate superconductor theory where the fermions are described by massless Dirac fermions \cite{a13,a14,a15,a16,a17,a18,a19}.

Dynamical chiral symmetry breaking (DCSB) occurs when the massless fermion acquires a nonzero mass through nonperturbative effects at low energy, but the Lagrangian keeps chiral symmetry when the fermion mass is zero. The Dyson-Schwinger equations (DSEs) provide a natural framework within which to explore DCSB and related phenomena. It is well known that in massless QED$_3$  DCSB occurs when the number of fermion flavors $N$ is less than a critical number $N_c$ \cite{a2,a3,a4,a5,a6}. Recently, by numerically solving the DSEs in the chiral symmetric phase of $QED_{3}$,  Fischer et al. display the anomalous dimension of the fermion vector dressing function in
the infrared domain for the case of bare vertex. They find that the wave-function renormalization has a power law behavior in the infrared region in Wigner phase while the fermion vector dressing function has no power law behavior in Nambu phase \cite{a11}.

It is shown that there exist antiferromagnetic correlations in the underdoped cuprates. In usual, one use the spin susceptibility to represent antiferromagnetic order. The theoretical calculations of the spin susceptibility are usually done in the framework of perturbation theory where nonperturbation effect is neglected. In this paper, we will study the staggered spin susceptibility in the framework of the Dyson-Schwinger approach and the effect of gauge boson mass on the staggered spin susceptibility.

\section{a model-independent integral formula for the staggered spin susceptibility}
For $N=2$, the spin operators with momenta near the momentum transfer $\vec{q}=(0,0)$, $(\pi,\pi)$, and $(\pi,0)$ have different forms when expressed in terms of $\psi_\alpha$. In Refs. \cite{a20,a21} the spin operator and the
corresponding spin correlation function are defined as following:
\begin{equation}
  S(x)=\frac12 \bar \psi_\alpha \Gamma \sigma_{\alpha\beta} \psi_\beta
\end{equation}
and
\begin{equation}
<S^{+}(q)S^{-}(-q)>=-\frac{1}{4}\int\frac{d^{3}p}{(2\pi)^{3}}Tr[\Gamma\mathcal{S}(p)\Gamma \mathcal{S}(p-q)],
\end{equation}
where
\begin{equation}
 \Gamma=\gamma^{0},~\textbf{1},~\left(
      \begin{array}{cc}
        0&1\\ 1&0\\
\end{array}
\right)
\end{equation}
 near
$\vec{q}=(0,0),(\pi,\pi),(\pi,0)$, respectively. The $\mathcal{S}(p)$ in Eq. (2) is fermion propagator of the QED$_3$ model in a large N expansion, which include leading order in the 1/N expansion.

At the mean-field level, the decay exponents of the three algebraic spin correlation functions are the same. When the Feynman diagrams for the spin susceptibility to leading order in the 1/N expansion are included, the decay exponents of the spin correlation near $\vec{q}=(0,0)$ and $\vec{q} =(\pi,0)$ do not change \cite{a20,a21}. Working beyond the mean-field level and including gauge fluctuations, Rantner and Wen calculated the nonzero leading
$O(1/N)$ order corrections to the staggered spin correlation function and obtain
\begin{equation}\label{2}
\mathcal{S}(p)=-iC\frac{p\cdot\gamma}{p^{2-\nu}},~\\
\\ \nu=\frac{32}{3N\pi^{2}},
\end{equation}
where $\nu$ is the nonzero anomalous dimension \cite{a20,a21,a22,a23}, which deduce the recovery to antiferromagnetic correlation at low energy.

Up to now, in all the above literature, the theoretical calculations of the spin susceptibility are usually done in the framework of perturbation theory where only the leading 1/N order corrections to the staggered spin correlation function are added to the mean field level. The primary goal of this paper is to derive a model-independent integral formula for the staggered spin susceptibility based on the linear response theory of the fermion propagator and then calculate it in the framework of the Dyson-Schwinger approach. In the past years, we studied the vacuum susceptibility which is an important parameter characterizing the non-perturbative properties of the QCD vacuum. By differentiating the dressed quark propagator with respect to the corresponding constant external field, the linear response of the nonperturbative dressed quark propagator to the constant external field can be obtained . Using this general method, we extract a rigorous and model-independent expression for the scalar, pseudoscalar, the vector, axial Vector, and the tensor vacuum Susceptibilities \cite{a24,a25,a26,b24,b25,b26,b27}. In this
paper, we shall take the same strategy to study the staggered spin susceptibility in QED$_{3}$.

The Lagrangian density of QED$_3$ with N flavors of massless fermion in Euclidean space reads:
\begin{equation}\label{eq1}
\mathcal{L}=\sum^N_{i=1}\bar\psi_i(\not\!\partial+\mathrm{i}e\not\!\not\!A)\psi_i+\frac{1}{4}F^2_{\rho\nu}+\frac{1}{2\xi}(\partial_\rho\ A_\rho)^2,
\end{equation}
where the $4 \times1$ spinor $\psi_i$ represents the fermion field, $i=1, \cdots, N$ are the flavor indices, and $\xi$ is the gauge parameter. In order to take into account the influence of the external field, we add an additional term $\Delta\mathcal {L}=-{\bar\psi}_{\gamma}(x)\tau^{+}_{\gamma\delta}\psi_{\delta}(x)\mathcal{V}(x)$ to the normal QED$_3$ Lagrangian, where $\tau^{+}=\tau_{1}+i\tau_{2}$ with $\tau_{i}$ being the Pauli matrices and
$\mathcal{V}(x)$ is a variable external field.

The fermion propagator $\mathcal{G}_{\alpha\beta}[\mathcal {V}](x)$ in the presence of the external field $\mathcal {V}$ can be written as
\begin{equation}\label{1}
   \mathcal{G}_{\alpha\beta}[\mathcal {V}](x)
    =\int\mathcal {D}\bar{\psi}\mathcal {D}\psi\mathcal
    {D}A\psi_{\alpha}(x)\bar{\psi}_{\beta}(0)\exp\{-\int d^{3}x[\mathcal {L}+\Delta\mathcal
    {L}]\},
\end{equation}
where the subscripts denote the flavor indices. If one assumes the external field $\mathcal {V}$ is weak and only considers the linear response term of $\mathcal{G}_{\alpha\beta}[\mathcal {V}](x)$, one has
\begin{eqnarray}\label{2}
    &&\mathcal{G}_{\alpha\beta}[\mathcal {V}](x) \nonumber \\
    &=&\int\mathcal {D}\bar{\psi}\mathcal {D}\psi\mathcal{D}A~\psi_{\alpha}(x)\bar{\psi}_{\beta}(0)\exp\{-S\}+\int\mathcal {D}\bar{\psi}\mathcal {D}\psi\mathcal{D}A\int d^{3}y[\psi_{\alpha}(x)\bar{\psi}_{\beta}(0)\bar{\psi}_{\gamma}(y)\tau^{+}_{\gamma\delta}\psi_{\delta}(y)\mathcal {V}(y)]\exp\{-S\} +\cdots \nonumber,\\
    &\equiv& G_{\alpha\beta}(x)+\mathcal {G}{^{\mathcal{V}}}_{\alpha\beta}(x)+\cdots
\end{eqnarray}
where $ G_{\alpha\beta}(x)=\langle 0| T\psi_{\alpha}(x)\bar{\psi}_{\beta}(0)| 0\rangle$ is the fermion propagator in the absence of the external field, $\mathcal {G}{^{\mathcal {V}}}_{\alpha\beta}(x)$ represents the linear response term of the fermion propagator
\begin{eqnarray}\label{2}
\mathcal {G}{^{\mathcal {V}}}_{\alpha\beta}(x)\equiv \langle 0| T\psi_{\alpha}(x)\bar{\psi}_{\beta}(0)|0\rangle_{\mathcal {V}}=\int d^{3}z \langle 0| T \psi_{\alpha}(x)\bar{\psi}_{\beta}(0)\bar{\psi}_{\gamma}(z)\tau^{+}_{\gamma\delta}\psi_{\delta}(z)|0\rangle \mathcal{V}(z).
\end{eqnarray}

Now we expand the inverse fermion propagator $\mathcal {G}^{-1}[\mathcal {V}]$ in powers of $\mathcal {V}$ as follows
\begin{eqnarray}\label{2}
    \mathcal{G}_{\alpha\beta}[\mathcal {V}](x)=G_{\alpha\beta}(x)-\int d^{3}y_{1} d^{3}y_{2} d^{3}zG_{\alpha\gamma}(x-y_{1})[\Gamma(y_{1},y_{2},z)]_{\gamma\delta}\mathcal {V}(z)G_{\delta\beta}(y_{2}) \nonumber\\
    =G_{\alpha\beta}(x)-\int d^{3}z \int \frac{d^{3}P}{(2\pi)^{3}}\int \frac{d^{3}q}{(2\pi)^{3}}e^{-i(q+\frac{P}{2})x} e^{iP\cdot z}G_{\alpha\gamma}(q+\frac{P}{2})[\Gamma(P,q)]_{\gamma\delta}\mathcal{V}(z) G_{\delta\beta}(q-\frac{P}{2}).
\end{eqnarray}
Setting $x=0$ in Eq. (9) and comparing it with the linear response term in Eq. (7), we obtain
\begin{equation}\label{10}
    \langle0 |T[{{\psi}_{\alpha}}(0)\bar{\psi}_{\beta}(0)\bar{\psi}_{\gamma}(z)\tau^{+}_{\gamma\delta}\psi_{\delta}(z)]| 0\rangle =-\int \frac{d^{3}P}{(2\pi)^{3}}\int \frac{d^{3}q}{(2\pi)^{3}}e^{iP\cdot z}G_{\alpha\gamma}(q+\frac{P}{2})[\Gamma_{P}(q,P)]_{\gamma\delta} G_{\delta\beta}(q-\frac{P}{2}).
\end{equation}
After multiplying $\tau^{-}_{\beta\alpha}$ on both sides of Eq. (10) and summing over the spinor and flavor indices, we obtain
\begin{eqnarray}\label{10}
&&\langle0 | T[\bar{\psi}_{\beta}(0)(\tau^{-})_{\beta\alpha}{\psi}_{\alpha}(0)\bar{\psi}_{\gamma}(z)(\tau^{+})_{\gamma\delta}\psi_{\delta}(z)]| 0\rangle\nonumber\\
&&=\int\frac{d^{3}P}{(2\pi)^{3}}\int\frac{d^{3}q}{(2\pi)^{3}}e^{iP\cdot z}Tr[G(q+\frac{P}{2})[\Gamma(P,q)] G(q-\frac{P}{2})\tau^-],
\end{eqnarray}
where the trace operation is over spinor and flavor indices. The staggered spin correlation function in momentum space is
\begin{equation}\label{3}
\langle S^{+}(P)S^{-}(-P)\rangle =\frac{1}{4}\int\frac{d^{3}q}{(2\pi)^{3}}Tr[G(q+\frac{P}{2})\Gamma(q,P) G(q-\frac{P}{2})\tau^-].
\end{equation}
where the spin density operator $S^{\pm}(x)=\frac{1}{2}\bar{\psi}(x)\tau^{\pm} \psi(x)$ with $\tau^{\pm}=\tau_{1}\pm i\tau_{2}$.

It is obvious that the staggered spin correlation depends on the precise form of the scalar vertex. In the ladder approximation, the scalar vertex satisfies the following Bethe-Salpeter equation
\begin{equation}\label{3}
\Gamma(q,P)=-\tau^+\otimes \textbf{1}-e^2\int\frac{d^{3}k}{(2\pi)^{3}}\gamma_{\mu}G(k+\frac{P}{2})\Gamma(k,P)G(k-\frac{P}{2})\gamma_{\nu} D_{\mu\nu}(k-q),
\end{equation}
where $\textbf{1}$ is the $4\times4$ unit matrix in spinor space. If one approximates the full scalar vertex with the bare one $\Gamma_{0}(q,P)=-\tau^+\otimes\textbf{1}$ , then our expression of the staggered spin susceptibility
reduces to the one given in Refs. [9,10]. It is obvious that the wave function renormalization $A(p^{2})$ is affected by the perturbative and nonperturbative effects of the scalar vertex.

Now we explicitly separate out the flavor part of the fermion propagator and the scalar vertex, i.e., $G(k) \rightarrow I_f\otimes G(k)$ and $\Gamma(q,P)\rightarrow -\tau^+ \otimes \Gamma(q,P)$, where $I_f$ is the unit matrix in flavor space. Then, using $Tr[\tau^{+}\tau^{-}]=4$, we obtain
\begin{equation}\label{3}
\langle S^{+}(P)S^{-}(-P)\rangle =-\int\frac{d^{3}q}{(2\pi)^{3}}Tr[\textbf{1}G(q+\frac{P}{2})\Gamma(q,P) G(q-\frac{P}{2})],
\end{equation}
where now the trace operation is over spinor indices and $G$ is the full fermions propagator. So far we have extracted a new expression for the staggered spin susceptibility. Here we note that this formula (14) is formally model-independent. However, the physical quantities which enter into it, such as the full fermions propagators and the vertices are usually obtained from QED$_3$ based models. Thus in practical calculations of the vacuum susceptibilities one usually resort to various models. For instance, as will be shown in detail below, in this paper we will calculate the staggered spin susceptibility within the framework of the BC1 vertex [6,11,12] approximation of the Dyson-Schwinger approach.

A diagrammatic representation of the staggered spin susceptibility is depicted in Fig. 1, where $\textbf{1}$ and $\Gamma$ are the bare and the full vertex, respectively. Here it is interesting to compare Eq. (14) with Eq. (2) given by Refs. \cite{a20,a21}. If one uses the bare scalar vertex approximation, i.e. $\Gamma=\textbf{1}$, Eq. (14) reduces into the staggered spin susceptibility given by Refs. \cite{a20,a21} (apart from the fact that the calculation of the fermions propagator $\mathcal{S}$ given by Refs. \cite{a20,a21} is quite different from the calculation of the full fermions propagator $G$ in the present work). Now it is clear that the nonperturbative vertex effects is neglected in previous paper.
\begin{figure}[htb]
\includegraphics[width=0.3\textwidth,clip]{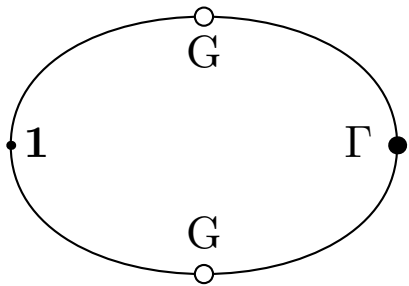}
\caption{ A diagrammatic representation of the staggered spin susceptibility }
\end{figure}

We now focus on the low-energy behavior of the staggered spin susceptibility. The dressed scalar vertex $\Gamma(q,P=0)$ and the staggered spin susceptibility has the following general form in the low energy limit:
\begin{equation}
\Gamma(q,P=0)=-\tau^+\otimes (F(q^{2})+i\gamma\cdot q H(q^{2})).
\end{equation}
and
\begin{equation}
 \langle S^{+}(0)S^{-}(0)\rangle=\int\frac{d^{3}q}{(2\pi)^{3}}\frac{F(q^{2})}{[q^{2}A^{2}(q^{2})+B^{2}(p^{2})]},
\end{equation}
where
\begin{equation}
F(q^{2})=1+2\int\frac{d^{3}k}{(2\pi)^{3}}\frac{F(k^{2})}{(k-q)^{2}(1+\Pi[(k-q)^{2}])[k^{2}A^{2}(k^{2})+B^{2}(k^{2})]}.
\end{equation}
The remaining task is then to calculate the staggered spin susceptibility in Wigner phase and Nambu phase. The wave-function renormalization $A(p^{2})$ in the above formula can be easily obtained by numerically solving the coupled DSEs for the fermion propagator.

\section{the effect of the gauge boson mass }
In Refs. \cite{a20,a21}, based on the so-called algebraic spin liquid picture, the authors analyzed the effect of the gauge boson mass acquired via the Anderson-Higgs mechanism on the staggered spin correlation. In order to make gauge field obtain a mass $\zeta$, we now introduce the additional interaction term between gauge field and complex scalar boson field
\begin{eqnarray}
\mathcal{L}_{B} = \sum^N_{i=1}[|(\partial_{\mu}+ieA_{\mu})\phi_{ i}|^{2}-\mu^{2}|\phi_{i}|^{2}-\lambda|\phi_{i}|^{4}]
\end{eqnarray}
which is so-called Abelian Higgs model or relative Ginzburg-Landau model \cite{KL}. The complex scalar field $\phi$ represents the bosonic holons, which has spin-$0$ and carry charge $e$. This Lagrangian describes the motion of the charge degrees of freedom of electrons on the CuO2 planes of underdoped cuprate superconductors. When $\mu^{2}> 0$, the system stays in the normal state and the vacuum expectation value of boson field $\langle \phi \rangle = 0$, so the Lagrangian respects the local gauge symmetry. When $\mu^{2}< 0$, the system enters the superconducting state and the boson field develops a finite expectation value $\langle \phi \rangle \neq 0$, then the local gauge symmetry is spontaneously broken and the gauge field acquires a finite mass $\zeta$ after absorbing the massless Goldstone boson. The finite gauge field mass is able to characterize the achievement of superconductivity. On the
other hand, the gauge mass obtains a mass via Anderson-Higgs mechanism implies that the gauge field is in confinement phase \cite{FR}, which deduce that the spions and holons are confined in superconducting phase (the spin-charge recombination). It is clear that the spinon and holon can not be observed in high-$T_{C}$ superconducting experiments, however, a well defined quasiparticle can be observed due to the spin-charge recombination in superconducting phase.

In QED$_{3}$ with Abelian Higgs model the gauge field couples to both the fermion field $\psi$ and the complex scalar boson field $\phi$. $\Pi_{\mu\nu}(q)=\Pi^{F}_{\mu\nu}(q)+\Pi^{B}_{\mu\nu}(q)$ is the total vacuum polarization tensor and the full inverse gauge boson propagator is
\begin{equation}\label{4}
D_{\mu\nu}^{-1}(q)=D_{\mu\nu}^{(0)-1}(q)+\Pi_{\mu\nu}(q),
\end{equation}
where $D_{\mu\nu}^{(0)-1}(q)$ is the free inverse gauge boson propagator, $\Pi_{F}(q)$ and $\Pi_{B}(q)$ are the polarization function from the fermion part and the boson part, respectively. The one-loop vacuum polarization $\Pi_{B}(q)$ has also been calculated by evaluating four Feynman diagrams \cite{KF,JH}. In the simplest approximation, Rantner and Wen take the following phenomenological form for the gauge propagator [22]:
\begin{equation}\label{5}
D_{\mu\nu}(q)=\frac{8}{N\sqrt{q^{2}+\zeta^{2}}}(\delta_{\mu\nu}-\frac{q_{\mu}q_{\nu}}{q^{2}}).
\end{equation}
Note that the gauge boson mass is added usually by hand in previous papers. In this paper, We will follow these papers to add the gauge mass by hand and study the influence of gauge boson mass on the staggered spin susceptibility in the framework of the Dyson-Schwinger approach. In this work, the gauge boson propagator in Landau gauge is given:
\begin{equation}\label{5}
D_{\mu\nu}(q)=\frac{1}{q^{2}[1+\Pi(q)]+\zeta^{2}}(\delta_{\mu\nu}-\frac{q_{\mu}q_{\nu}}{q^{2}})
\end{equation}
and we choose the BC1 vertex ansatz $\Gamma_\nu(p,k)=\frac{1}{2}[A(p^2)+A(k^2)]\gamma_\nu$ \cite{a6,a11,b1}. Thus in the Landau gauge the coupled DSEs with gauge boson mass $\zeta$ is obtained:
\begin{equation}\label{eq10}
A(p^{2})=1+\frac{1}{p^{2}}\int\frac{\mathrm{d}^3k}{(2\pi)^3}\frac{A(p^{2})+A(k^{2})}{A^{2}(k^{2})k^{2}+B(p^{2})}\frac{A(k^{2})(p\cdot q)(k\cdot q)/q^{2} }{[q^{2}(1+\Pi(q^{2}))+\zeta^{2}]},
\end{equation}
\begin{equation}\label{eq11}
B(p^2)=\int\frac{\mathrm{d}^3k}{(2\pi)^3}\frac{B(k^2)[A(p^2)+A(k^2)]}{[A^2(k^2)k^2+B^2(k^2)][q^2(1+\Pi(q^2))+\zeta^{2}]},
\end{equation}
\begin{equation}\label{eq12}
\Pi(q^2)=N\int \frac{\mathrm{d}^3k}{(2\pi)^3}\frac{A(k^2)A(p^2)[A({p}^2)+A(k^2)]}{q^2[A^2(k^2)k^2+B^2(k^2)]}\frac{[2k^2-4k\cdot q-6(k\cdot q)^2/q^2]}{[A^2(p^2)p^2+B^2(p^2)]},
\end{equation}
where $q=p-k$. Here we want to stress that the $B(p^2)$ in Eq. (23) has two qualitatively distinct solutions. The “Nambu” solution, for which $B(p^2)\neq 0$, describes a phase in which (a) chiral symmetry is dynamically broken, because one has a nonzero fermion mass function, and (b) the dressed fermions are confined, because the propagator described by these functions does not have a Lehmann representation. The alternative “Wigner” solution, for which $B(p^2)\equiv 0$ describes a phase in which chiral symmetry is not broken and the dressed fermions are not confined. In addition, it should be noted that BC1 vertex Ansatz violates the fundamental QED Ward identity, namely $q^{\mu} \Gamma_{\mu}=G^{-1}(p)-G^{-1}(k)$. If the  Ward identity is not satisfied, then the transversality of the photon self-energy is also compromised ($q^{\mu} \Pi_{\mu\nu}(q)$ ought to vanish, but it does not). A better Ansatz is given in Refs. \cite{BC,ap}. However, this particular pathology is not easy to detect at the level of Eq. (24), because we have suppressed the tensorial structure of the photon self-energy, keeping only the scalar form factor $\Pi(q^2)$. Furthermore, the BC1 vertex has the advantage that the equations are simplified significantly and it already contains main qualitative features of the solution employing the BC vertex in the infrared region, as was demonstrated by the numerical calculations given in Refs. [6,12]. This is the main reason why we still choose the BC1 vertex in our work.

It is well known that one can obtain two types of solution by iterating the above coupled DSEs, the Nambu solution and the Wigner solution. After solving the above coupled DSEs by means of iteration method, we can numerically calculate the integration in Eq. (16) in both Nambu phase and Wigner phase. Because of the importance of anomalous dimension exponent for the staggered spin susceptibility \cite{a20,a21,a22,a23}, we firstly study the momentum dependence of $A(p^{2})$ in the infrared region for several  gauge boson mass for N=2. In Fig. 2, the dependence of $A(p^{2})$ on the momentum for several values of the gauge boson mass are shown. From the obtained numerical results one finds that in Wigner phase $A(p^{2})$ enhances when the gauge boson mass monotonically increases, while in Nambu phase it decreases with the increase of the gauge boson mass. It is well known that when the gauge boson mass is zero, the vector dressing function $A(p^{2})$ in Wigner phase has a power law behavior in the infrared region \cite{a11}. From Fig. 2, It is clear that the anomalous dimension exponent will change for several gauge
boson mass since these curves are not parallel in infrared region. However, the authors of Refs. \cite{a20,a21} still use the same anomalous dimension exponent $\nu$ to study the staggered spin susceptibility when the gauge boson mass is nonzero.

\begin{figure}[htb]
\includegraphics[width=0.45\textwidth,clip]{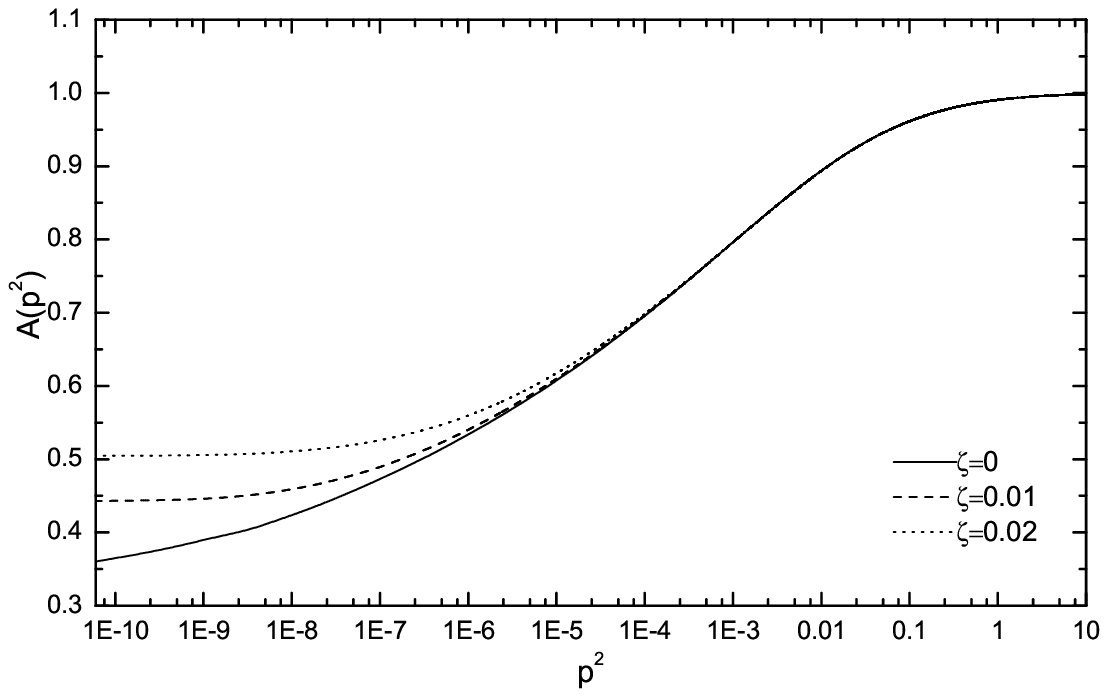}
\includegraphics[width=0.45\textwidth,clip]{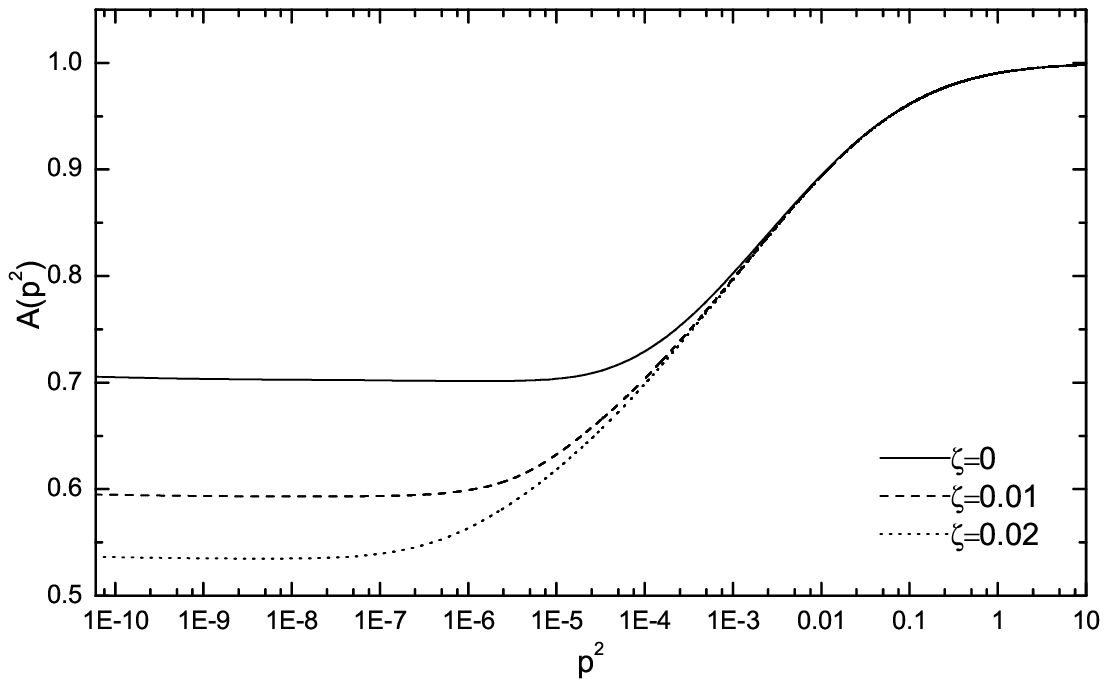}
\caption{The dependence of  $A(p^{2})$ on the momentum for several gauge boson mass $\zeta$ in Wigner phase and Nambu phase }
\end{figure}

On the other hand, the difference between $A(p^{2})$ in Wigner phase and Nambu phase is more and more smaller as the gauge boson mass increases from Fig. 2. When the gauge boson mass reaches a critical value $\zeta=0.024$, the dependence of $A(p^{2})$ on the momentum in these two phases become the same, as is shown in Fig. 3. In fact, Nambu phase disappears when $\zeta=0.024$.

\begin{figure}[htb]
\includegraphics[width=0.7\textwidth,clip]{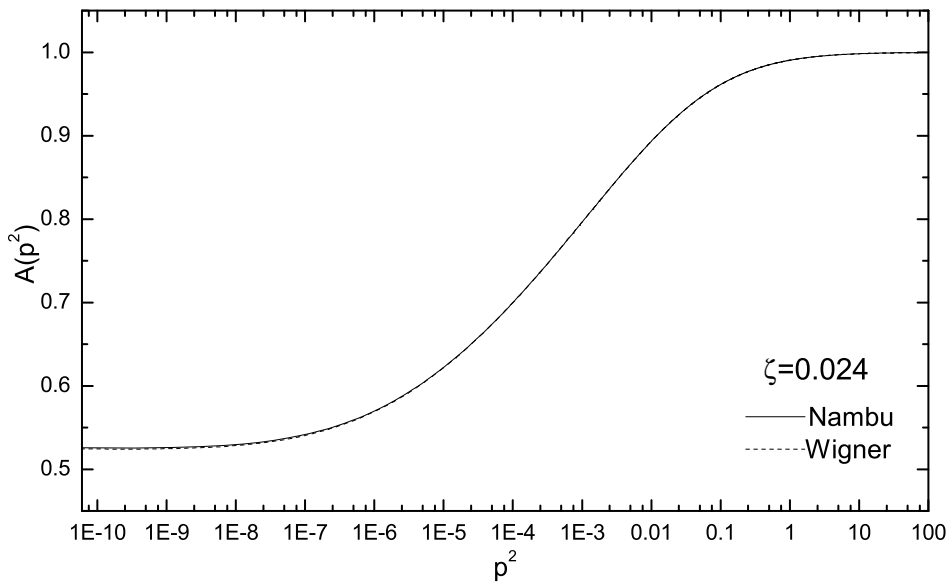}
\caption{The dependence of  $A(p^{2})$ on the momentum  at $\zeta$ = 0.024 in Wigner phase and Nambu phase }
\end{figure}

Now let us to study qualitatively the influence of the gauge boson mass on the staggered spin susceptibility from the competition between the antiferromagnetic order and the superconducting order. The staggered spin susceptibility is used to represent the antiferromagnetic order in QED$_3$ model. On the other hand, the gauge boson mass is proportional to the superfluid density, so the gauge boson mass can be used to describe the superconducting order. Due to the competition between the antiferromagnetic order and the superconducting order in high temperature cuprate superconductors, it is obvious that the opening of a gap in the gauge fluctuations
will spoil the antiferromagnetic correlation.

\begin{figure}[htb]
\includegraphics[width=0.7\textwidth]{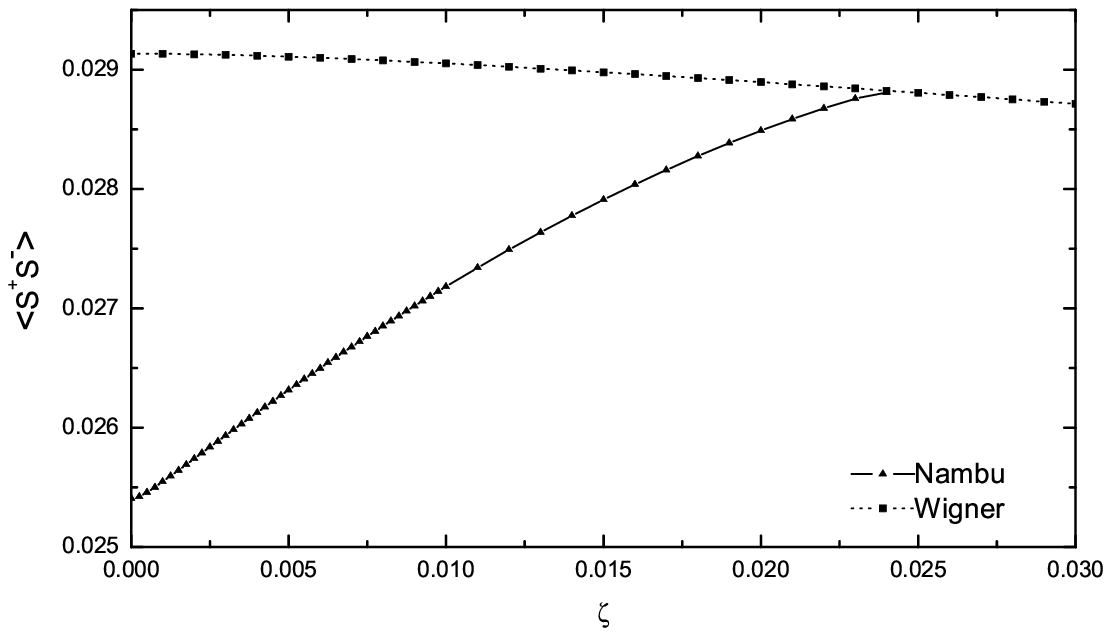}
\caption{The dependence of $<S^{+}S^{-}>$ in the low energy limit for several gauge boson masses}
\end{figure}

From the large momentum behavior of $A(p^{2})$, $ B(p^{2})$, and $F(p^{2})$, we see that the staggered spin susceptibility given by Eq. (16) is linearly divergent, and this divergence cannot be eliminated through the standard renormalization procedure. In order to extract something meaningful from the staggered spin correlation, one needs to subtract the linear divergence of the free staggered spin susceptibility from Eq. (16), which is analogous to the regularization procedure for calculating the chiral susceptibility in QCD, which is quadratically divergent (see for example, Ref. \cite{he}). We define the regularized staggered spin susceptibility by
\begin{equation}\label{3}
\langle S^{+}(0)S^{-}(0)\rangle _{R} =\langle S^{+}(0)S^{-}(0)\rangle - \langle S^{+}(0)S^{-}(0)\rangle_{ free},
\end{equation}
where the free staggered spin susceptibility $\langle S^{+}(0)S^{-}(0)\rangle_{ free}$ is calculated at the mean field level.

The numerical result for  the staggered spin susceptibility for $F(p^{2})=1$ case is given in Fig. 4.  We find that the gauge boson mass suppresses the staggered spin susceptibility in the low energy limit in Wigner phase. On the contrary, the staggered spin susceptibility increase with the gauge boson mass increasing in Nambu phase. When the value of the gauge boson mass reaches 0.024, the staggered spin susceptibility takes the same value in Nambu
phase and Wigner phase. Once the gauge boson mass exceeds this critical value, Nambu phase disappears. Here it should be noted that the imaginary part of staggered spin susceptibility is related to the scattering function which can be detected by experiment \cite{ex}, Therefore, in order to compare the staggered spin susceptibility with the related experiment, one should continue it into real frequencies (more detail can be found in Ref. \cite{a21}), we will discuss this question in the near future.

Now we focus on the staggered spin susceptibility in Winger phase. Experimentally it has been proved that the staggered spin correlations decrease with the increase of the doping $x$. Rantner and Wen have explained the unusual property based on the algebraic spin liquid plus the spin-charge recombination picture \cite{a20,a21}. In fact, this strange experimental behavior can be explained naturally in our paper. At zero temperature, the superfluid density in the underdoping region depends on the doping $x$ as $\rho=x/a$, where $a$ is the lattice spacing \cite{a27,a28}. As the doping $x$ increases, the superfluid density increases. Since the gauge boson mass is proportional to the superfluid density, it also increases when the doping $x$ increases. That is to say, the staggered susceptibility decreases with the increase of the doping. On the other hand, it is shown that the gauge boson mass suppresses the staggered spin susceptibility in the low energy limit in Fig. 4. Our results in Wigner phase give a qualitative physical picture on the competition and coexistence between the antiferromagnetic order (the staggered spin susceptibility) and the superconducting orders (the gauge boson mass ) in high temperature cuprate superconductors. On the contrary, the staggered spin susceptibility increase with the gauge boson mass increasing in Nambu phase. The conclusion in Winger phase fail to show the true physical picture.

\section{conclusions}
The primary goal of this paper is to investigate the effect of the gauge boson mass on the staggered spin susceptibility. Based on the linear response theory of the fermion propagator in the presence of an external scalar field, we first derive a model-independent integral formula, which expresses the staggered spin susceptibility in terms of objects of the basic quantum field theory: dressed propagator and vertex. When one approximates the scalar vertex
function by the bare one, this expression, which includes the influence of the nonperturbative dressing effects, reduces to the expression for the staggered spin susceptibility obtained using perturbation theory in previous works. Then we calculate numerically the staggered spin susceptibility in both Nambu phase and Wigner phase when the gauge boson acquire a mass. It is found that when the gauge boson mass increases, the staggered spin susceptibility in Wigner phase decreases, while the staggered spin susceptibility in Nambu phase increases. When the gauge boson mass reaches a critical value, Nambu phase disappears. In addition, in Winger phase, we also find
that the superconducting order suppresses the antiferromagnetic order. Our result may help to explain why in high-temperature superconducting experiments the antiferromagnetic order decreases with the increase of the doping.

\bigskip
\section{acknowledgement}
This work is supported in part by the National Natural Science Foundation of China under Grants No. 11347212, No. 11275097, No. 11475085, and No. 11105029 and by the Natural Science Foundation of Jiangsu Province under Grant No. BK20130387 and the Fundamental Research Funds for the Central Universities (under Grant No. 2242014R30011).

\end{document}